\documentstyle[sprocl,epsfig]{article}

\input{psfig}

\newcounter{figureno}                         

\def\Journal#1#2#3#4{{#1} {\bf #2}, #3 (#4)}

\def\NPB{{\em Nucl. Phys.} B}
\def\PLB{{\em Phys. Lett.}  B}
\def\PRL{\em Phys. Rev. Lett.}
\def\PRD{{\em Phys. Rev.} D}
\def\ZPC{{\em Z. Phys.} C}
\def\EPJC{{\em Eur. Phys. J.} C}

\def\ra{\rightarrow}

\def\be{\begin{equation}}
\def\ee{\end{equation}}
\def\bea{\begin{eqnarray}}
\def\eea{\end{eqnarray}}

\begin{document}

\title{MEASUREMENTS OF $ B \rightarrow X_s \gamma$ AND STUDY OF $B \rightarrow X_s l^+ l^-$}

\author{G. EIGEN}

\address{Department of Physics, University of Bergen, Allegaten 55,
5007 Bergen, \\ Norway \\ E-mail: eigen@asfys2.fi.uib.no}

\maketitle\abstracts{The updated CLEO measurement of the electromagnetic penguin
process $B \ra X_s \gamma$ is presented and 
compared to the ALEPH result. Implications on new physics are discussed and the status
of recent searches for $B \ra X_s l^+ l^-$ modes is given.
}

\section{Introduction}

\begin{figure}[h]
\vskip -3.0cm
\begin{center}
\setlength{\unitlength}{1cm}
\begin{picture}(10,2.7)
\put(0.,-0.0)
{\mbox{\epsfysize=1.9cm\epsffile{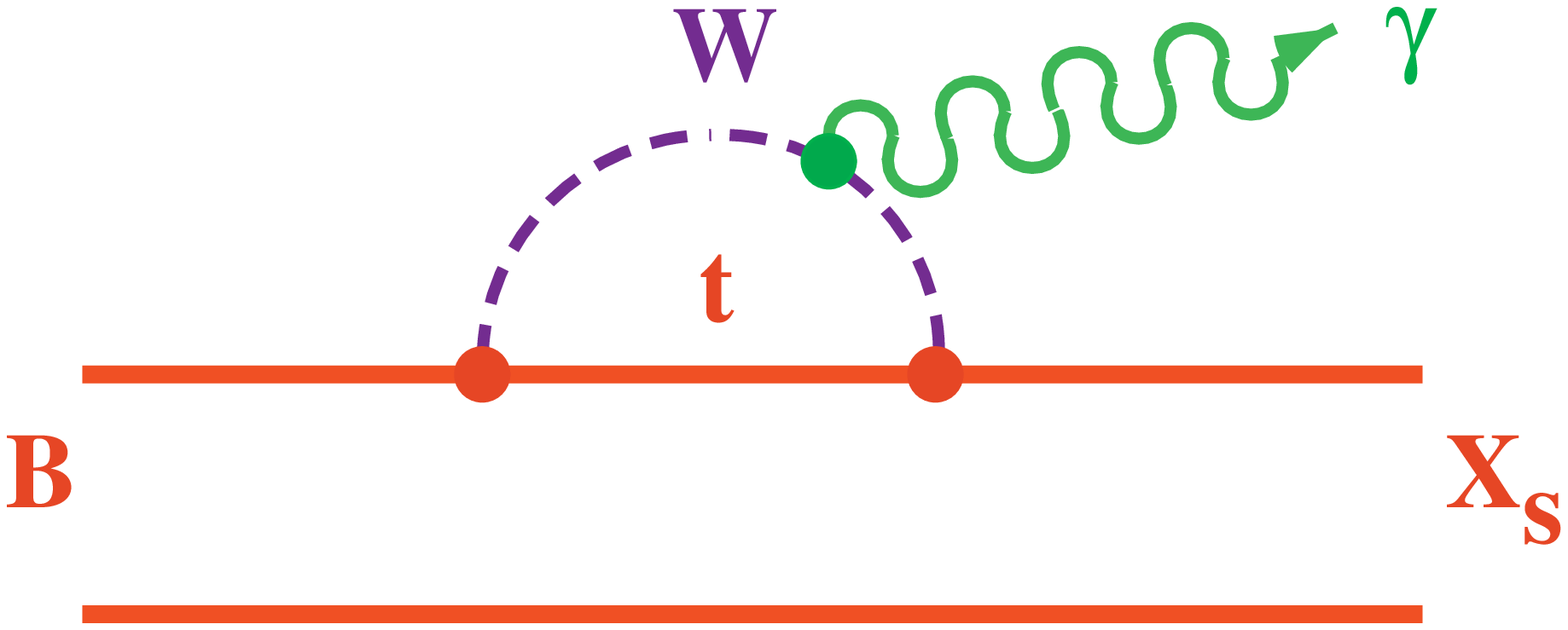}}\hspace*{5mm}\vspace{0.2 cm}
\mbox{\epsfysize=2.5cm\epsffile{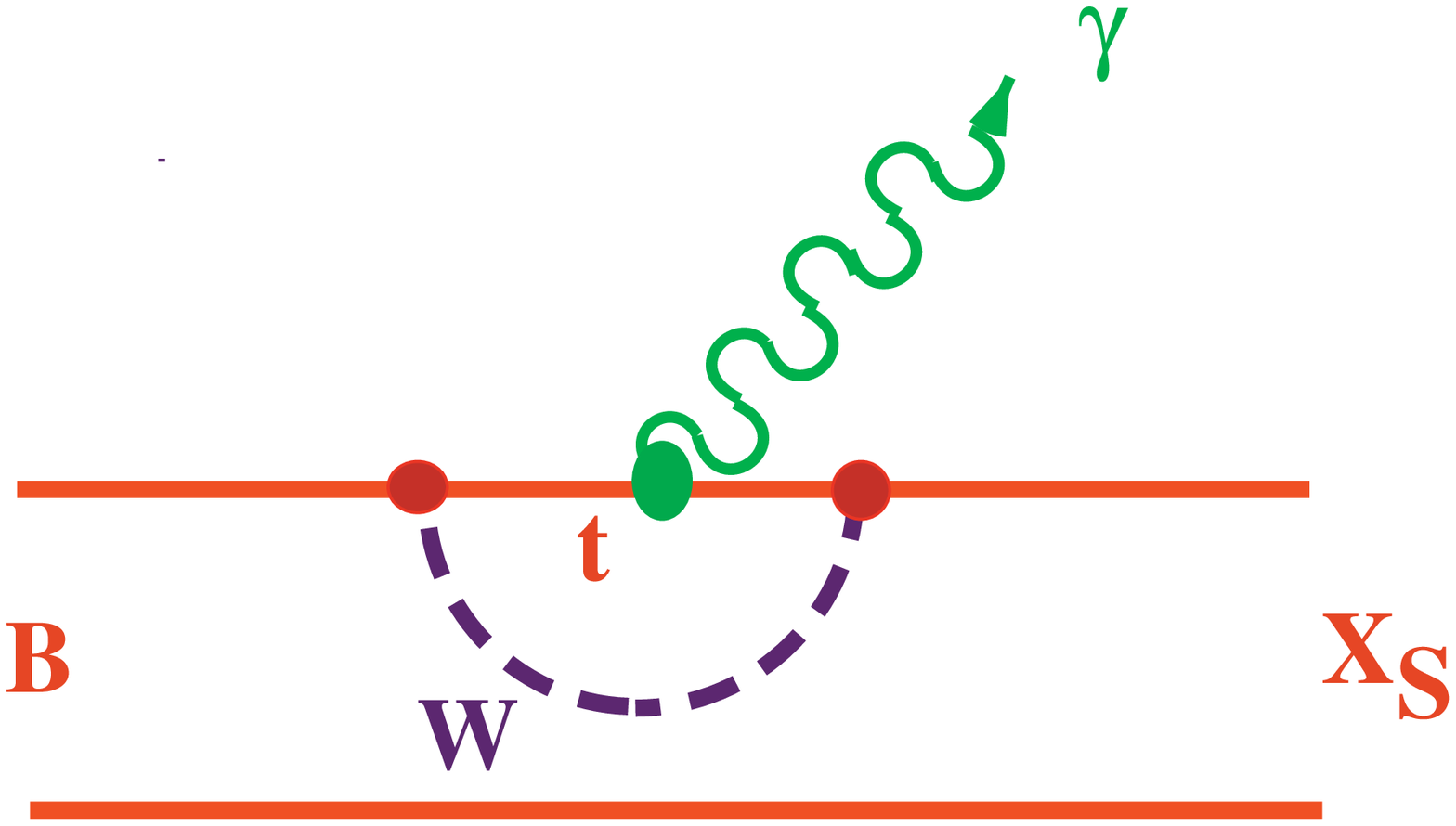}}}  
\end{picture}
\caption{Lowest-order Feynman diagrams for $B \ra X_s \gamma$ in the SM. 
\label{fig:bsg1}}
\end{center}
\end{figure}

Flavor-changing neutral currents (FCNC) are forbidden in the Standard Model (SM) at
tree level. They are, however, induced at higher orders via penguin processes or box diagrams.
One such process is $ B \ra X_s \gamma$, which is mediated by electromagnetic (em) penguin loops. 
The lowest-order SM diagrams are shown 
in Figure~\ref{fig:bsg1}. The dominant contribution arises from the magnetic dipole operator 
${\cal O}_7$. QCD corrections, however, cause mixing with the other  
operators $ {\cal O}_1$ to ${\cal O}_6$ and ${\cal O}_8$. The dominant short-distance effects 
are separated from the non-perturbative long-distance parts
into perturbatively-calculable renormalization-scale-dependent Wilson coefficients. 
In $B \ra X_s \gamma$ all perturbative contributions 
can be absorbed into a single effective function, $ \tilde C^{eff}_7 (\mu) $, which reduces to
$ C^{(0)eff}_7 (5.0) = -0.300$ in leading-order QCD.\cite{bu}
However, a $25\%$ uncertainty in the decay rate arises from
varying the renormalization scale between $m_b/2$ and $2 m_b$. In
next-to-leading order (NLO) calculated recently \cite{ch}
this uncertainty is reduced by a factor 1.5-2.\cite{BABAR}

In the Standard Model the inclusive $B \ra X_s \gamma$ decay rate is \cite{ch,bu2}
\begin{equation} 
\Gamma (B \ra X_s \gamma) = {m_b^5 \over 32 \pi^4} G_F^2 \alpha_{em} \mid V_{tb}
V_{ts}^\ast \mid ^2 ( \mid \tilde C^{eff}_7 (\mu) \mid ^2 + A ),
\end{equation}

\noindent
where $m_b$, $G_F$, $\alpha_{em} $, $A$, and $V_{tb} V_{ts}^\ast $ denote the $b$-quark mass, the
Fermi constant, the em coupling constant, the bremsstrahlung contribution and
Cabibbo-Kobayashi-Maskawa (CKM) matrix elements for $t \ra b$ and $ t \ra s $ couplings,
respectively. The $ B \ra X_s 
\gamma $ branching fraction is calculated by scaling the semileptonic decay rate, yielding 
$ B(B \ra X_s \gamma) = (3.28 \pm 0.33) \times 10^{-4} $ in NLO QCD.\cite{ch}
Since the branching fraction is sensitive to additional contributions from  $e.g.$ 
a charged Higgs boson, a new gauge boson or a charged supersymmetric (SUSY) particle in the loop, 
$ B \ra X_s \gamma$ is well-suited to probe new physics. A branching fraction measurement 
$e.g.$ can be used to constrain the SUSY parameter space.  
CLEO was the first experiment to observe this process both in the exclusive $K^\ast \gamma$ final
state \cite{am} and the inclusive channel.\cite{al}

\section{The recent CLEO $ B \ra X_s \gamma$ Measurement}

The inclusive  $ B \ra X_s \gamma$ analysis has been updated recently by CLEO  
using a sample of $3.3 \times 10^6 B \bar B$ pairs. The photon energy, which is nearly half the
$B$ mass ($ < E_\gamma > \simeq 2.4$~GeV), 
is the highest observed in any $B$ decay. The Fermi motion of the $b$-quark 
inside the B mesons and the B meson momentum in the lab frame yield a  
Doppler-broadened photon line.
According to spectator model predictions, 85\% - 94\% of the photons are in the
$2.1$ GeV $\leq E_\gamma \leq 2.7$~GeV energy range.\cite{ali}
While backgrounds from B decays are small, those from continuum processes, such as
$ e^+ e^- \ra q \bar q \gamma $ and  $ e^+ e^- \ra q \bar q  $, are significant. To minimize
these backgrounds an event shape analysis is combined with exclusive final-state reconstruction,
considering only high-energy photons as candidates that are inconsistent with originating 
from a $ \pi^0$ or  $ \eta $ decay.

At the $\Upsilon (4S)$ the event shape of hadronic $B \bar B$ decays is spherical, while 
that of $q \bar q$ continuum is back-to-back. In  $ B \ra X_s \gamma$ events the spherical
symmetry is distorted by the hard photon. The event shape resembles  
more that of $ q \bar q$ continuum with initial-state radiation (ISR).
The ratio of second-to-zeroth Fox-Wolfram moments $R_2$ and the sphericity $S_\perp$   
are used to separate the signal from $ q \bar q$ continuum.  
To suppress ISR continuum background 2 variables are evaluated
in the rest frame of the $e^+ e^-$  following the radiation of a hard photon:
the ratio of Fox Wolfram moments $ R^\prime_2$, where the photon is excluded, and 
$ \cos \theta^\prime$, where $ \theta^\prime $ is 
the angle between the photon and the thrust axis of the rest of the event.
In addition, the energy flow inside $20^o$ and $30^o$ cones 
parallel and antiparallel with the photon direction are calculated. 
With help of a neural network the 8 variables are combined into
a single shape variable $r_{sh}$, which favors values of $ r_{sh} \approx +1$ for
$B \ra X_s \gamma$ signal and $ r_{sh} \approx -1$ for continuum background.

\begin{figure}[h]
\vskip -3.0cm
\begin{center}
\setlength{\unitlength}{1cm}
\begin{picture}(15,6.0)
\put(-0.2,-0.0)
{\mbox{\epsfysize=6cm\epsffile{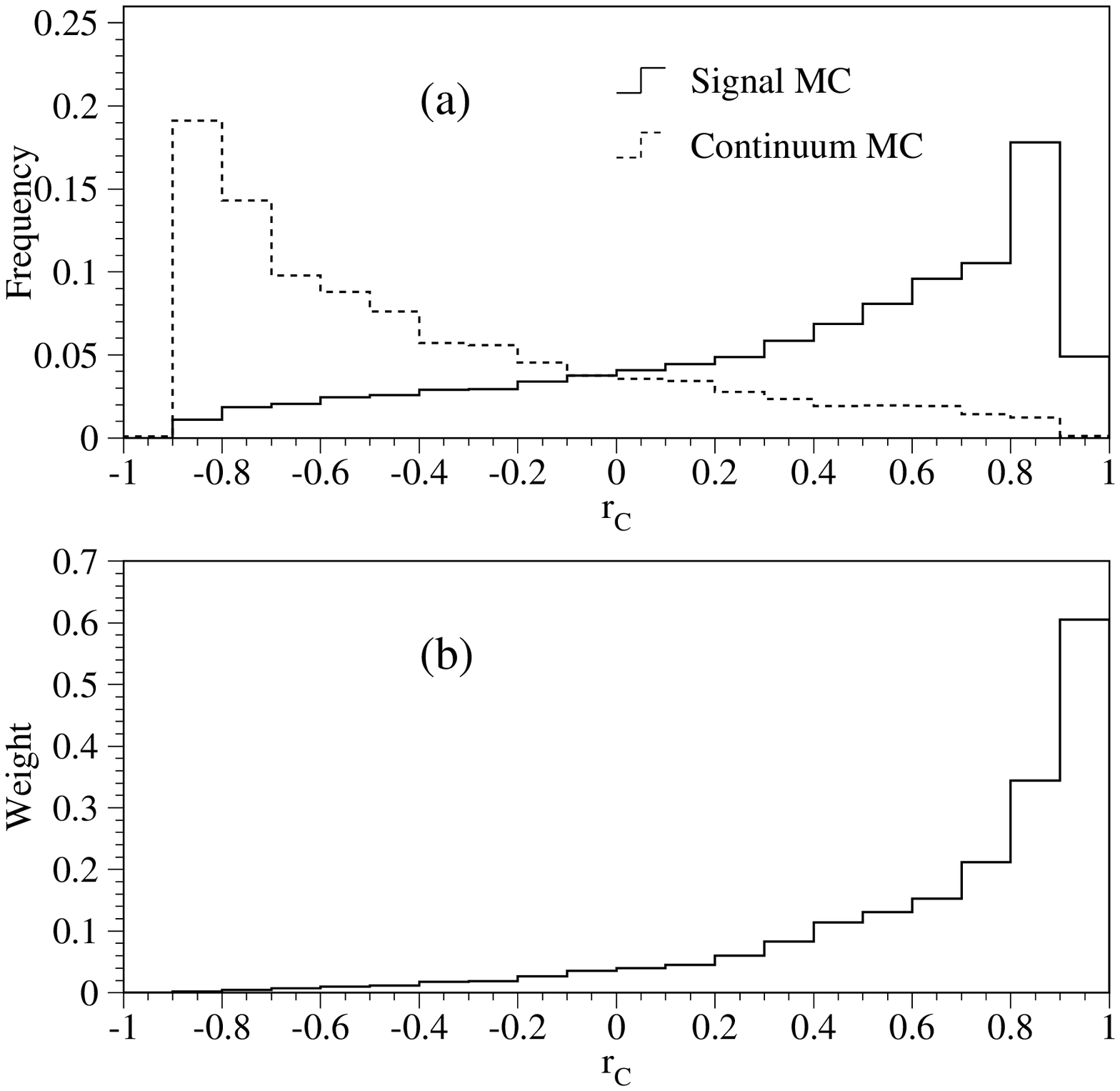}}\hspace*{-1mm}
\mbox{\epsfysize=6cm\epsffile{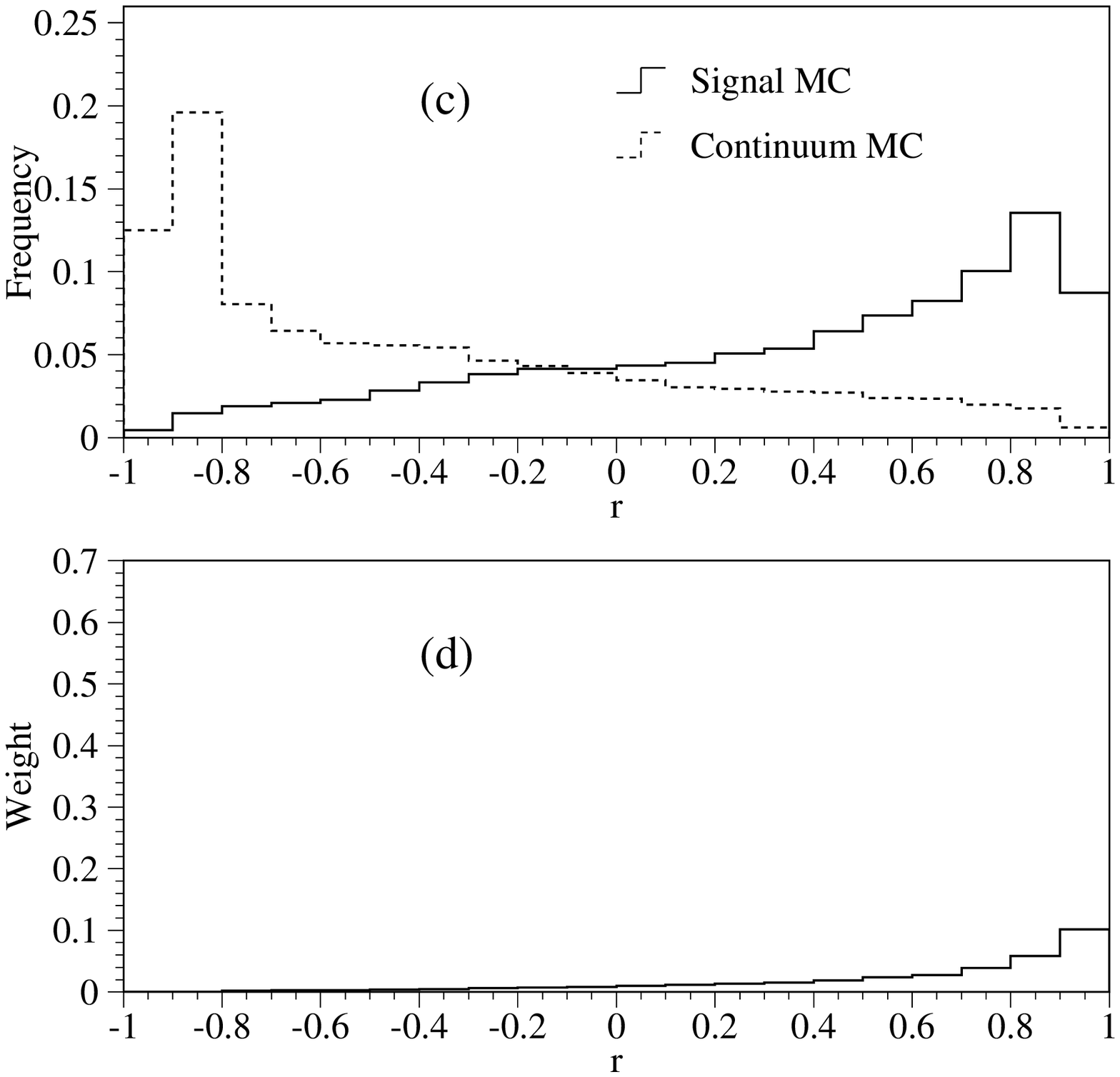}}}  
\end{picture}
\caption{a) Distributions in the shape variable $r_c$ for MC samples of
$B \ra X_s \gamma$ and $q \bar q$ continuum for events with $\chi^2_B < 20$; b) 
event weighting function for $r_c$; c) corresponding distributions in shape variable $r_{sh}$
for events failing reconstruction; d) event weighting function for $r_{sh}$.
\label{fig:rc}}
\end{center}
\end{figure}

For additional suppression of continuum background the $X_s$ system is reconstructed in 
exclusive final states. Considered are modes containing either a $K^\pm$ or a 
$K^0_S \ra \pi^+ \pi^-$ plus 1-4 $\pi$'s with at most one $\pi^0$ in addition to the hard photon. 
A global $\chi^2$ based on the beam-constrained mass $M_B$, the energy difference 
$\Delta E$, the $dE / dx$ of the charged particles as well as the $K^0_S$ and $\pi^0$ masses 
is used for candidate selection. For the
combination with the lowest $\chi^2$ a second shape variable $r_c$ is constructed using a neural 
network. The input variables are $r_{sh}$, the $\chi^2_B$ value obtained from 
$M_B$ and $\Delta E$ alone and  $\mid \cos \theta_{tt} \mid$, where $\theta_{tt}$ is
the angle between the thrust axes of the candidate $B$ and the 
rest of the event. Again signal events peak at $r_c \approx +1$, while continuum background 
peaks at $r_c \approx -1$. Figures~\ref{fig:rc}a,c show the $r_{c}$ and $r_{sh}$
distributions for signal and $q \bar q$ continuum obtained from Monte Carlo (MC).

To optimize the signal-to-background ratio, the candidates are weighted by
$w = s/ [ s + (1+\alpha)b]$,
where $s$ and $b$ are the expected signal and background yields for a given value of
$r_c$ and $\alpha$ is the luminosity scale factor between the 
$\Upsilon(4S)$ and the off-resonance data samples.
Since just $\sim 20\%$ of signal events
are exclusively reconstructed, a weighting based on $r_{sh}$ alone is used for 
selected events which fail the exclusive reconstruction ($\chi^2_B > 20$). The event weighting 
functions are displayed in Figures~\ref{fig:rc}b,d.

Backgrounds from $B \bar B $ decays, where the photon originates from a $\pi^0$
or an $\eta$ decay, are estimated from Monte Carlo. However, the $\pi^0$ and $\eta$
momentum spectra are adjusted to those observed in the data. The spectra are obtained by 
treating the $\pi^0$ or $\eta$ like a photon, performing the same analysis as discussed
above for photons using weights $w$. This procedure helps to account for inaccuracies 
in the $ b \ra c  W^-$ and  $ b \ra u  W^-$ event generation and for omissions of modes
such as $B \ra X_s g$. Further analysis details are given in reference [9].

\begin{figure}[h]
\begin{center}
\setlength{\unitlength}{1cm}
\begin{picture}(14,5.8)
\put(-0.1,-0.1)
{\mbox{\epsfysize=6.3cm\epsffile{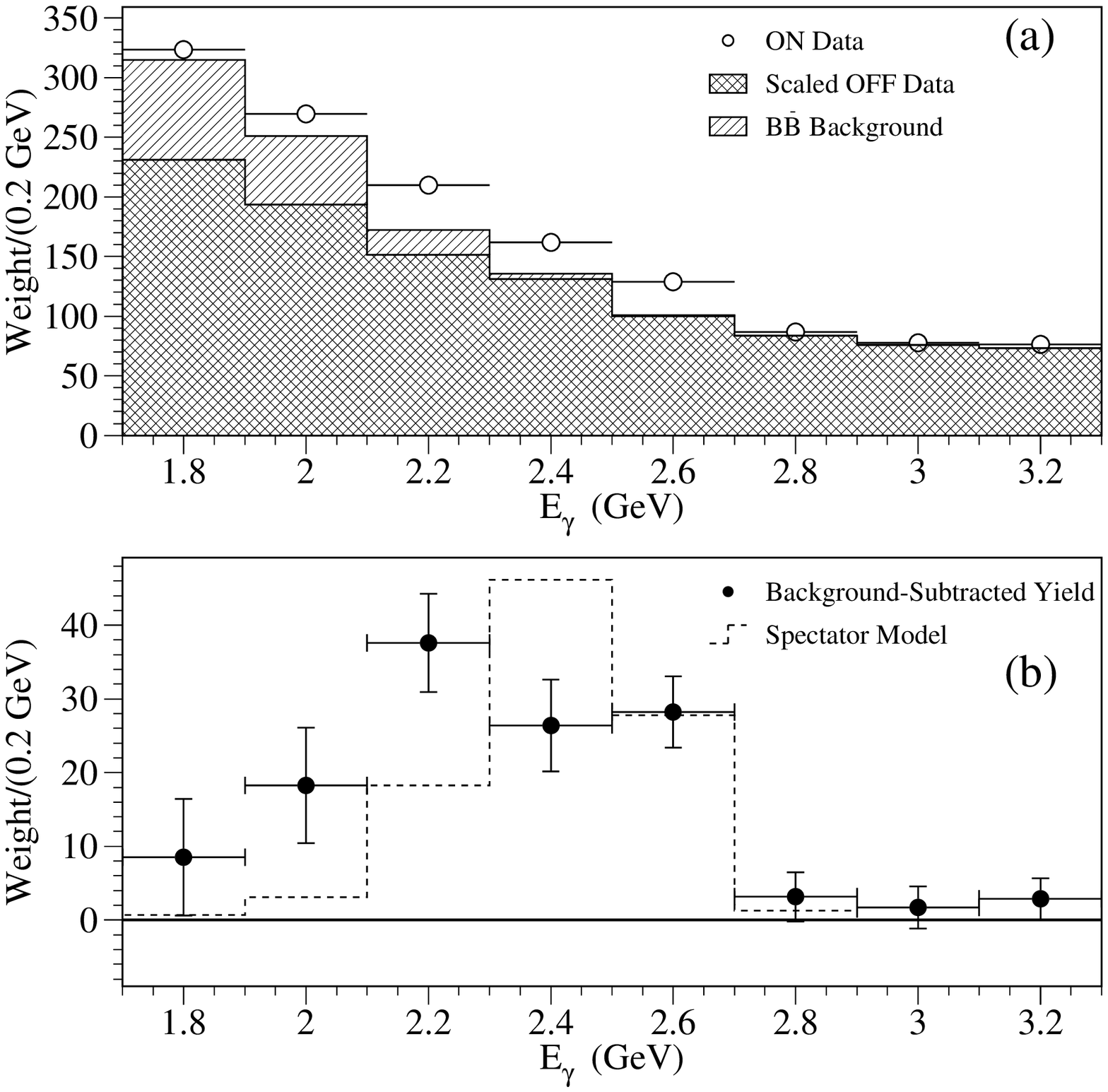}}\hspace*{-3mm}
\mbox{\epsfysize=5.8cm\epsffile{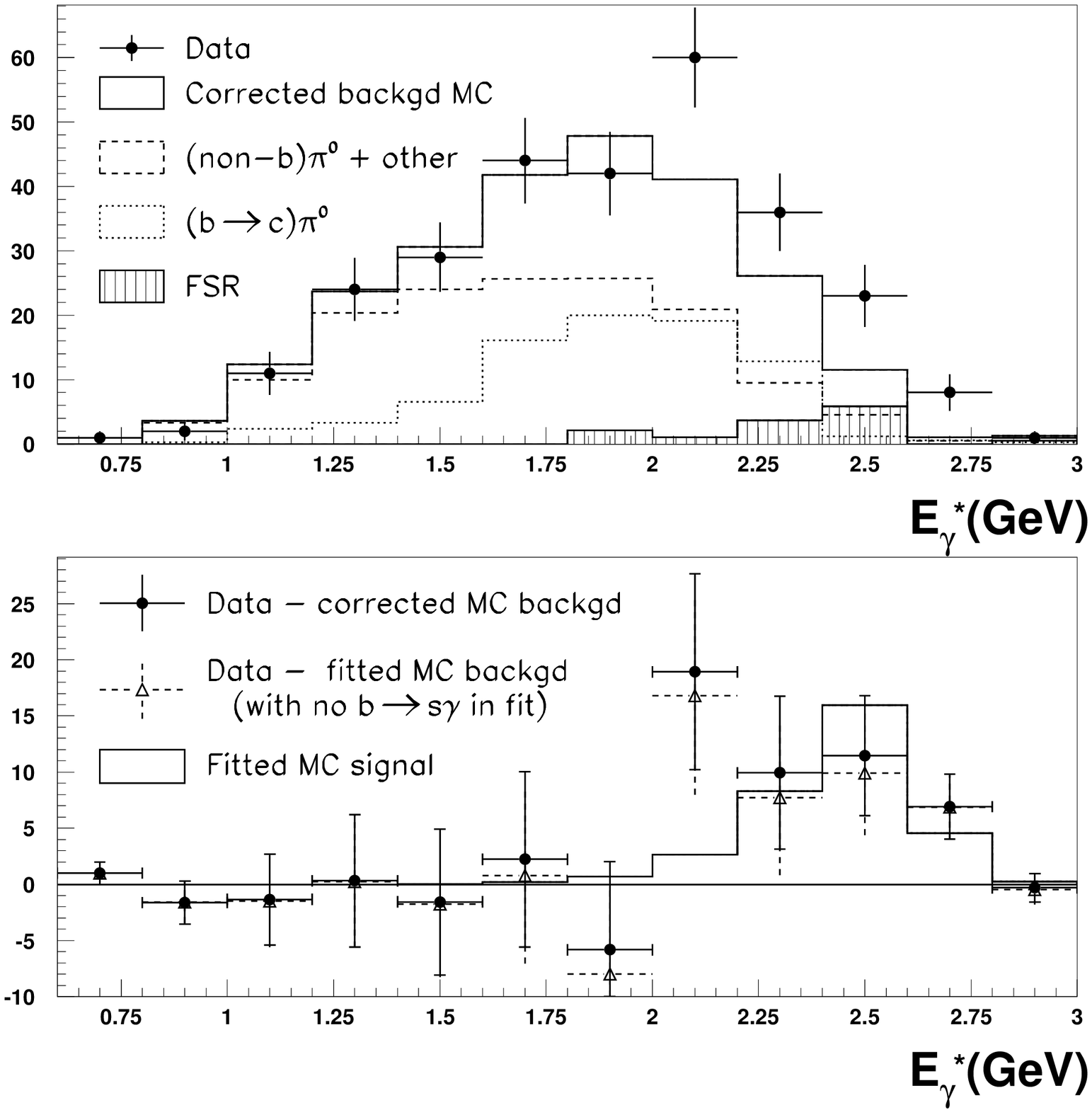}}}  
\end{picture}
\caption{a) Observed weighted photon energy spectrum and background contributions from CLEO; b)
background-subtracted photon yield and predicted shape of $B \ra X_s \gamma$
signal.
\label{fig:cleo}}\vspace{-0.2cm}
\caption{top) Observed photon energy spectrum and background 
contributions from ALEPH; bottom) background-subtracted photon yield from  
multivariate fits. \label{fig:aleph}}
\end{center}
\end{figure}

Figure~\ref{fig:cleo}a shows the weighted photon spectrum and the estimated  
contributions from $ B \bar B$ and continuum backgrounds. The background-subtracted photon 
spectrum plotted in Figure~\ref{fig:cleo}b shows a clear signal in the 2.1-2.7~GeV region.
The observed shape is consistent with the  
spectator model prediction.\cite{ali}
The weighted event yields in two bins below and one bin above the signal region
are consistent with the background parametrization. The background-subtracted 
weighted event yield in the 2.1-2.7~GeV window amounts to 
$92.21 \pm 10.26_{stat} \pm 6.46_{sys}$ events. The systematic error consists of
a $\pm1\%$ uncertainty in the $ q \bar q$ continuum subtraction and a $20\%$ uncertainty
in the $B \bar B$ subtraction, both added in quadrature.

The weighted efficiency in the 2.1-2.7~GeV is determined from the Ali-Greub
model,\cite{ali} which includes gluon bremsstrahlung and higher-order radiative effects.
The Fermi momentum of the $b$-quark and the mass of the spectator quark are varied
such that on average $< m_b > = 4.88 \pm 0.1$~GeV. For the
hadronization process both the JETSET algorithm \cite{sj} and a mix of $K^*$ resonances
are used such that the $X_s$ mass distribution predicted in the Ali-Greub model \cite{ali}
is reproduced. A weighted efficiency of $\epsilon_w = (4.70 \pm 0.23 \pm 0.23 \pm
0.04 \pm 0.33) \times 10^{-2}$ is determined in the signal region. The 4 errors
represent uncertainties due to spectator model inputs, 
hadronization of the $X_s$ system, the production ratio of
$B^+ B^-$ and $B^0 \bar B^0$ and detector modelling, respectively.

Combining the  background-subtracted weighted event yield with the \break weighted 
efficiency results in a branching fraction of 
\begin{equation}
 B(B \ra X_s \gamma) =(3.15 \pm 0.35_{stat} \pm 0.32_{sys} \pm 0.26_{model}) 
\times 10^{-4}.
\end{equation}
This measurement is in good agreement with the Standard Model prediction.
It supercedes the previous CLEO result, since  
60\% additional data are included and improved analysis techniques are used.
The $95\% $ confidence level (CL) limits including all systematic effects are 
\begin{equation}
2.0 \times 10^{-4} < B(B \ra X_s \gamma) < 4.5 \times 10^{-4}.
\end{equation}

\section{The $ B \ra X_s \gamma$ Result from ALEPH}

At LEP the $b$ and $\bar b$ quark are separated into opposite hemispheres.
The $B \ra X_s \gamma $ hemisphere has low multiplicity, a single displaced
vertex and a high-energy photon, whereas the opposite hemisphere contains a
typical $b$ hadron decay that is used for a $b$-tagging. 
In a sample of $8.9 \times 10^5 b \bar b$ pairs, ALEPH selects
$B \ra X_s \gamma $ candidates with an inclusive reconstruction algorithm. 
First, probabilities $P$ to originate from a $B \ra X_s \gamma $ decay
are defined for the candidate photon and the other particles
(charged tracks, $\pi^0$'s, $K^0_S$'s and $K^0_L$'s) in the candidate jet, using
rapidity, momentum and impact parameter. Second, the candidate
photon and up to 8 particles in decreasing order of probability $P$ are added until an 
invariant mass of the system closest to the $B$ mass is obtained. 
In addition to standard $b$ hadron selection criteria, the candidate photon has to
be inconsistent with a $\pi^0$ decay; the $\gamma \gamma$ invariant mass
when combined with another photon has to lie above 0.2~GeV. The photon energy has to 
exceed 10~GeV and $\theta^\ast_\gamma$, the angle between the photon in the
rest frame of the jet and the jet direction has to satisfy
$ \cos \theta^\ast_\gamma  < 0.55$. While the signal is uniformly 
distributed, the background peaks at $\cos \theta^\ast_\gamma = +1$. 
The mass of the candidate $B$ has to be within 0.7~GeV of the nominal $B$ mass and 
the mass of the $X_s$ system has to be below 4~GeV.

The resulting sample is devided into 8 subsamples, using the major axis of the
electromagnetic shower ellipse $\sigma_l$ to separate merged $\pi^0$'s 
from photons, the jet energy $E_{jet}$, and the probability $P^{opp}_{hem}$ 
that all charged tracks in the opposite hemisphere are
consistent with originating from the primary vertex. This procedure 
provides an effective  separation of signal photons from backgrounds that originate from
$\pi^0$'s, final state radiation (FSR) and other sources.
The signal is enhanced in the subsample with $ E_{jet} > 32$~GeV, 
$\sigma_l < 2.3$~cm and $-\log P^{opp}_{hem} > 2.2$. The photon energy $E^\ast_\gamma$ in the
jet rest frame is plotted for this subsample in Figure~\ref{fig:aleph}. 
An excess is seen in the 2.2-2.8~GeV region. A binned log-likelihood 
fit is performed in all eight subsamples to extract the number of signal events 
and background events due to FSR, $\pi^0$'s from $b$ decays, $\pi^0$'s from 
non-$b$ decays and other sources. Of the 1560 candidates
$ 69.4 \pm 19.7$ are found to be signal events.
The largest backgrounds result from $\pi^0$ decays. 
The efficiency for selecting $B \ra X_s \gamma$ signal 
is $\epsilon = 12.8 \pm 0.3 \%$.
Further details are discussed in the publication.\cite{ba}

From the yield in the 2.2-2.8~GeV energy interval one obtains
an inclusive branching fraction of
\begin{equation}
B(B \ra X_s \gamma) =(3.11 \pm 0.8 \pm 0.72) \times 10^{-4}.
\end{equation}
This is in good agreement with the CLEO result and the SM prediction.
The largest systematic error contributions are due to background shapes (0.462),
background MC statistics (0.376), background MC composition (0.185), 
and energy calibration uncertainties (0.182). 

\section{ Implications on New Physics}

\begin{figure}[h]
\vskip -3.0cm
\begin{center}
\setlength{\unitlength}{1cm}
\begin{picture}(9,3.0)
\put(0.,-0.0)
{\mbox{\epsfysize=2.7cm\epsffile{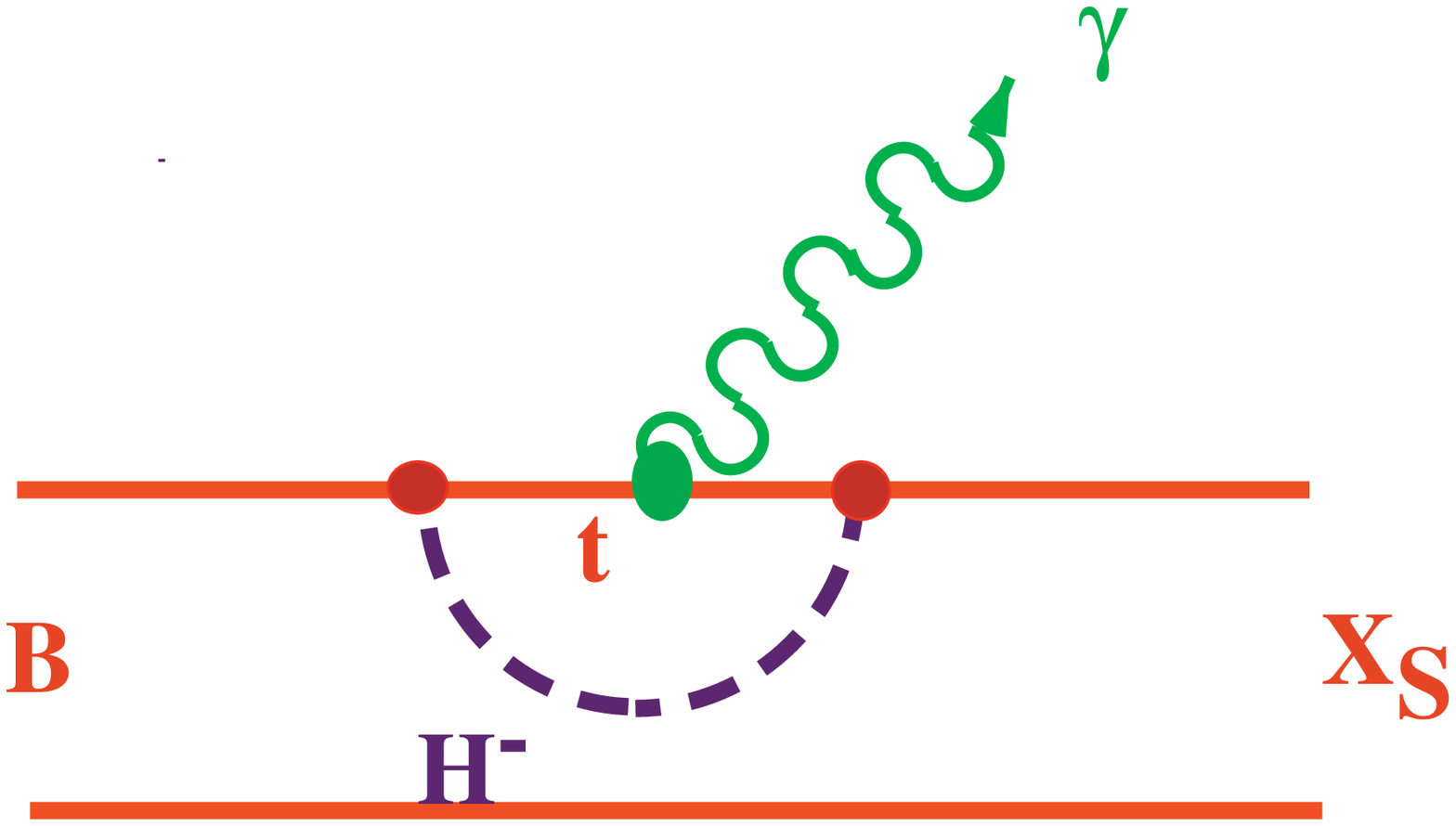}}\hspace*{5mm}
\mbox{\epsfysize=2.7cm\epsffile{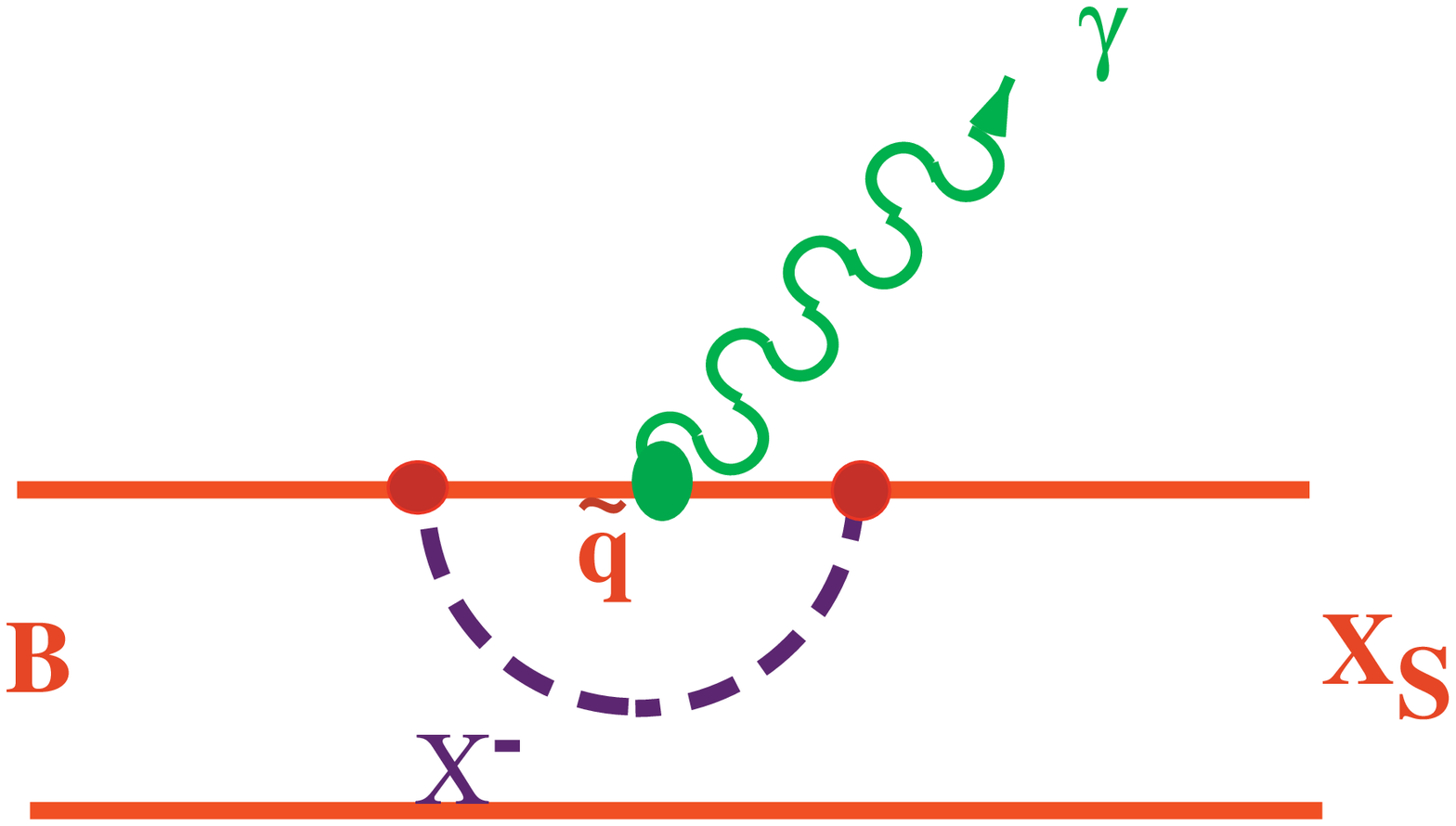}}}  
\end{picture}
\caption{Feynman diagrams for new physics contribution to $B \ra X_s \gamma$. 
\label{fig:bsg3}}
\end{center}
\end{figure}

\begin{figure}[h]
\begin{center}
\setlength{\unitlength}{1cm}
\begin{picture}(14,5.0)
\put(-0.6,-0.5)
{\mbox{\epsfysize=5cm\epsffile{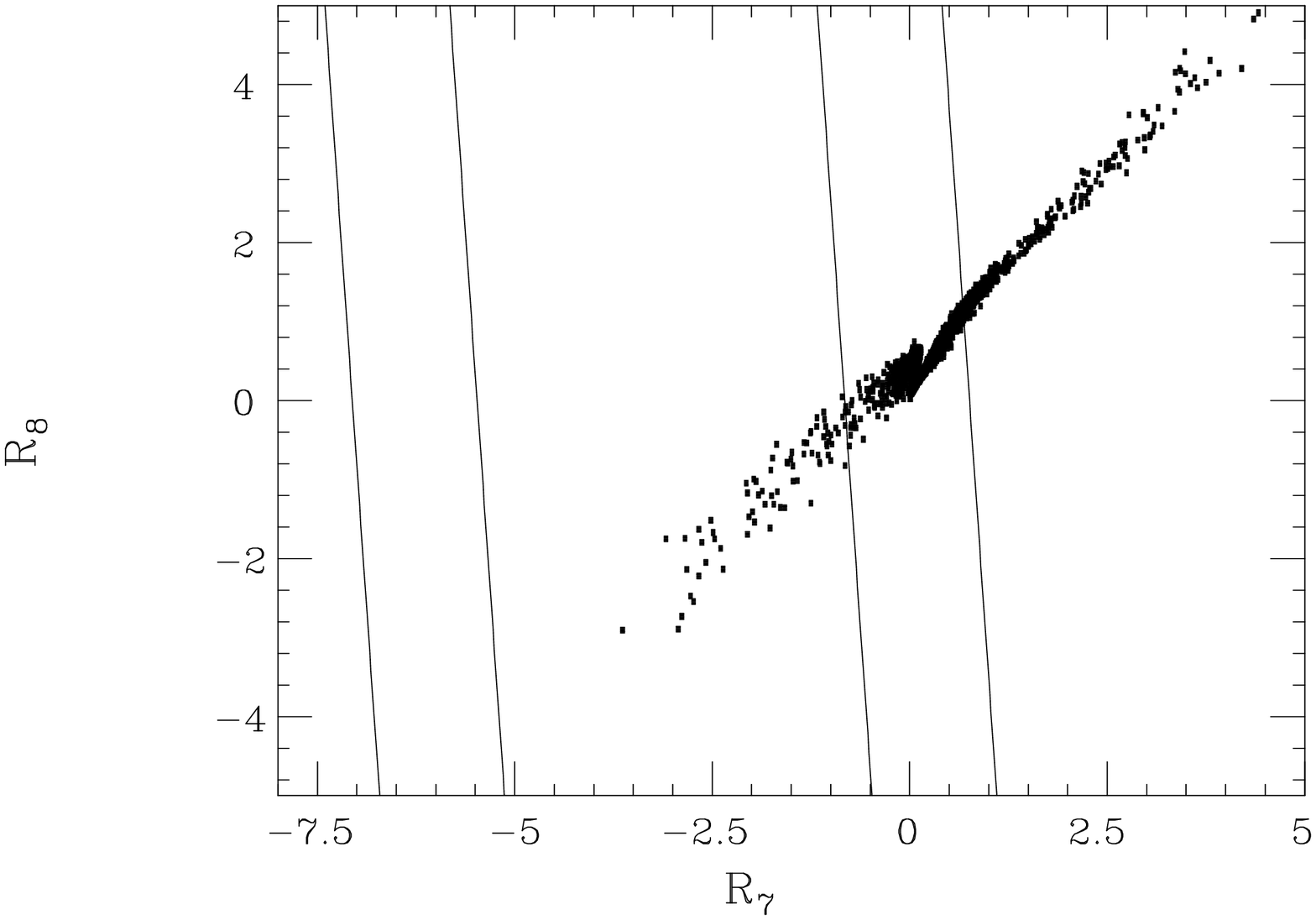}}\hspace*{-6mm}
\mbox{\epsfysize=5cm\epsffile{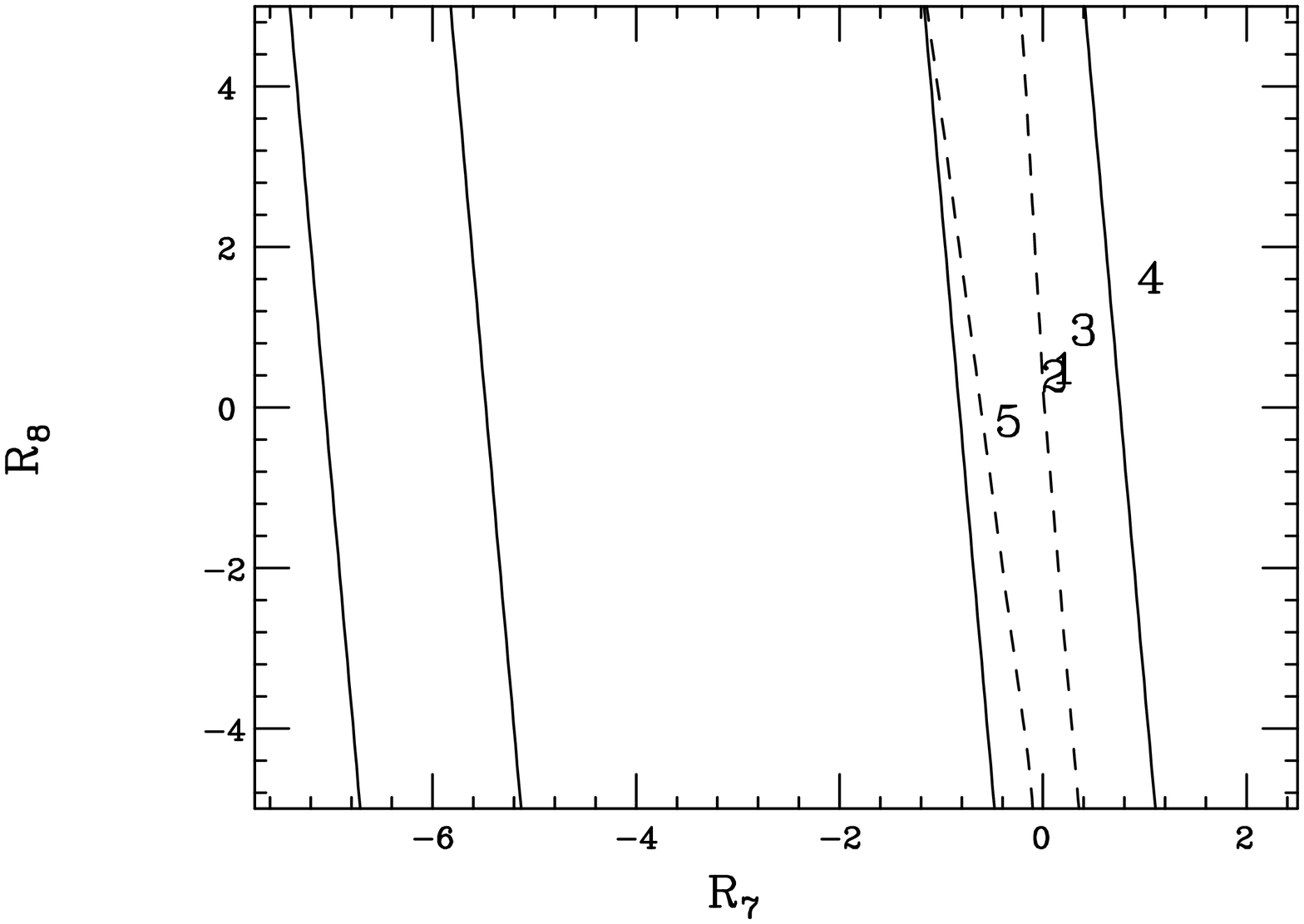}}}  
\end{picture}
\caption{Scatter plot of $R_8$ vs $R_7$ for a) solutions in the SUGRA model and b) for 5 NLC SUGRA
points. The allowed region from the CLEO measurement lies inside the 2 sets of solid diagonal bands. 
The dashed lines indicate a potential 10\% measurement.
\label{fig:susy}} 
\end{center}
\end{figure}

New physics processes can enhance the $B \ra X_s \gamma $ decay rate. For
example, in SUSY models one expects contributions from a charged
Higgs and charginos in the loop as shown in Figure~\ref{fig:bsg3}. These yield 
new contributions $C_7^{new}(M_W)$ and $C_8^{new}(M_W)$
to the SM Wilson coefficients at the $M_W$-scale.  
Due to operator mixing,
$B \ra X_s \gamma$ then limits the possible values for $C_i^{new}(M_W)$.
To study the effect of $B \ra X_s \gamma$ on SUSY,
solutions in the SUSY parameter space have been generated 
using the minimal supergravity model (SUGRA).\cite{he}
The ranges of the input parameters are $ 0 < m_0 < 500$~GeV, $50 < m_{1/2} <
250$~GeV, $ -3 < A_0/m_0 < 3$, $ 2 < \tan \beta < 50$ and $ m_t = 175$~GeV.\cite{he}
Each solution is only kept if it is not in violation with present SLC/LEP
constraints and Tevatron direct sparticle production limits. For these
solutions the ratios $R_7= C_7^{new}(M_W)/C^{SM}_7(M_W)$ and 
$R_8= C_8^{new}(M_W)/^{SM}_8(M_W)$ are calculated. The resulting scatter plot
in the $R_7-R_8$ plane is shown in Figure~\ref{fig:susy}a. Many solutions are already
in conflict with the present $B \ra X_s \gamma$ measurement from CLEO (region inside solid
bands).

To demonstrate the reach of $B \ra X_s \gamma$ in probing SUSY in comparison to 
that of high-energy colliders, 5 points in the SUGRA
model chosen at Snowmass 1996 for an NLC study are examined.\cite{da} Point~3 is the common point  
for a comparison of SUSY studies at NLC, LHC and upgraded Tevatron. 
The sparticle mass spectra for these 5 points are obtained by SUGRA relations
and their contribution to  $B \ra X_s \gamma$ can be computed.
Figure~\ref{fig:susy}b shows the results in the $R_7$-$R_8$ plane.\cite{he2} 
Also shown are the constraints obtained from fits to the present CLEO data (solid line)
and a future $10\%$ measurement assuming a SM value for the branching 
fraction (dashed line). While point 4 is already excluded by the present CLEO measurement,
all but point 5 lie outside the dashed diagonal band.
Thus, rare $B$ decays provide a complementary approach to high-energy colliders in searching for SUSY.

\begin{figure}[h]
\vskip -3.0cm
\begin{center}
\setlength{\unitlength}{1cm}
\begin{picture}(9,2.7)
\put(0.,-0.0)
{\mbox{\epsfysize=2.0cm\epsffile{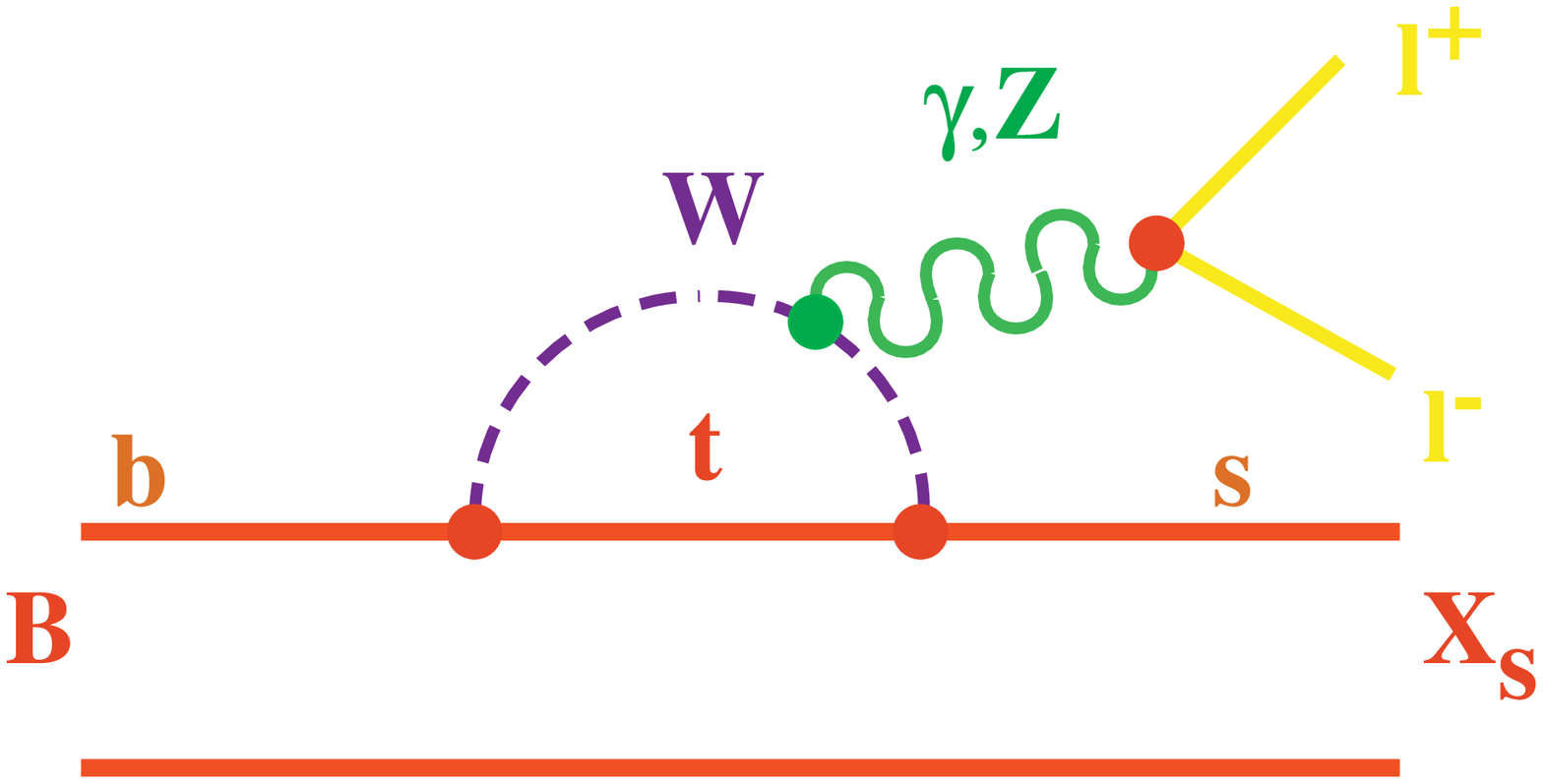}}\hspace*{5mm}\vspace*{-2.4cm}
\mbox{\epsfysize=2.5cm\epsffile{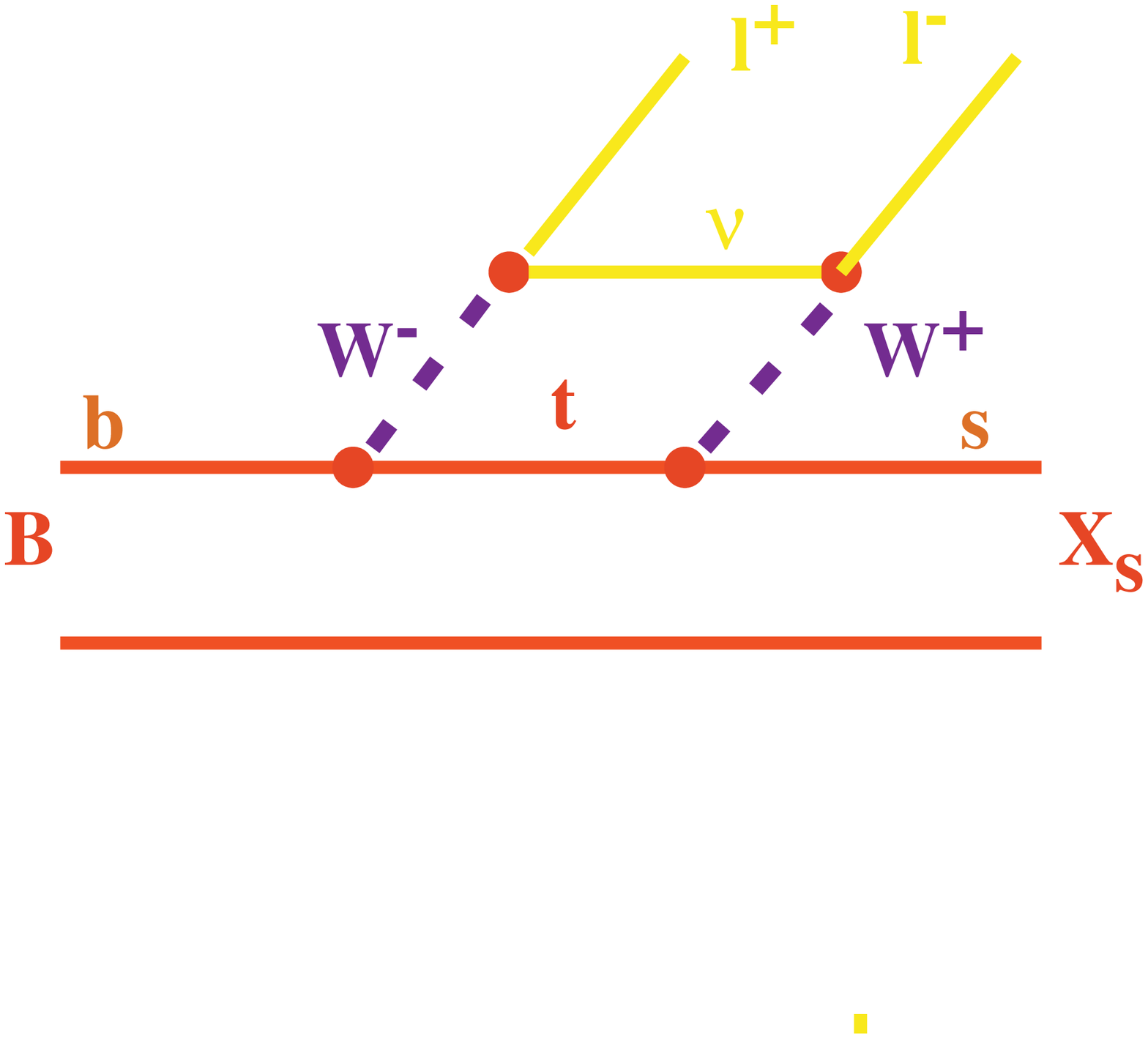}}}  
\end{picture}
\caption{Lowest order Feynman diagrams for $B \ra X_s l^+ l^-$ in SM. 
\label{fig:bsll1}}
\end{center}
\end{figure}

\section{ Study of $ B \ra X_s l^+ l^-$ decays at CLEO}

Another interesting and related process is $ B \ra X_s l^+ l^-$, where the photon
is replaced with a lepton pair. The lowest-order SM diagrams are
shown in Figure~\ref{fig:bsll1}. In addition to the magnetic penguin operator ${\cal O}_7$, the 
electroweak operators ${\cal O}_9$ and ${\cal O}_{10}$ contribute, yielding 3
scale-dependent Wilson coefficients: $C_7(\mu)$, $C_9(\mu)$ and $C_{10}(\mu)$.
Due to QCD effects, which are smaller here than in $B \ra X_s \gamma$, 
again operator mixing occurs. Their contributions are absorbed into the
effective Wilson coefficients $ C_7^{(0)eff}(\mu)$ and $ C_9^{eff}(\mu)$.
The decay rate has been calculated in NLO. 
However, to maintain a scheme-independent result,  
only  $ C_9(\mu)$ is calculated in NLO while  
LO is retained for all other Wilson coefficients.
New physics can affect any of these 3 Wilson coefficients. 
Due to the additional $\alpha_{em}$ coupling, the branching fractions predicted in SM
are almost 2 orders of magnitude smaller than those for corresponding
$B \ra X_s \gamma$ modes.

Using a sample of $3.3 \times 10^6$ $B \bar B$ pairs CLEO has searched for both
inclusive $X_s l^+l^-$ channels and 8 exclusive final states, where a $K$ or 
$K^\ast$ recoils against $\mu^+ \mu^-$ or $ e^+ e^-$. A detailed discussion of
the inclusive analysis is given in reference [16]. In the exclusive analysis
a Fisher discriminant method is used 
in addition to standard $B$ selection criteria 
to suppress continuum background. 
It is constructed from the distributions of thrust and sphericity
of the $B$ candidate, the $B$ direction in the lab frame and the angle
between the thrust axes of the candidate $B$ and the rest of the event. 
The candidate $B$ is selected by a $\chi^2$ method, which includes $\Delta E$, the masses 
of intermediate resonances, and both the $ dE/dx$
and time-of-flight information of all charged tracks in the final state.
The combination with the lowest 
$\chi^2$ is retained. Since double semileptonic decays are
a dominant background source, the missing energy of the event is constrained to $< 20\%$. 
To remove contributions from the color-suppressed $ B \ra \psi^{(\prime)} K^{(\ast)}$ modes,
events with dilepton masses near the $\psi$ and $\psi^\prime$ mass peaks are excluded.
Signal regions are defined in the $M_B$ - $\Delta E$

\begin{figure}[h]
\begin{center}
\setlength{\unitlength}{1cm}
\begin{picture}(7,8.5)
\put(-0.5,-0.3)
{\mbox{\epsfysize=8.8cm\epsffile{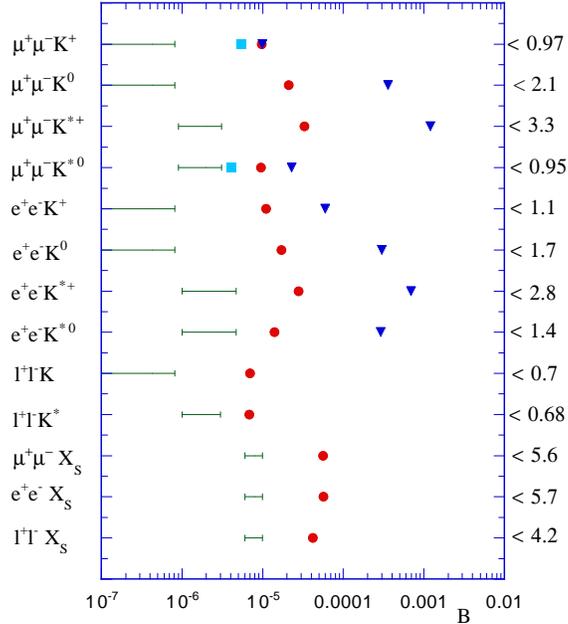}}}  
\end{picture}
\caption{Branching fraction upper limits of $B \ra X_s l^+ l^-$ modes at $90\%$ CL from
CLEO (solid points) in comparison to a range of SM predictions (solid lines), 
recent results from CDF (squares) and previous upper limits (triangles). The numbers on the 
right-hand side show the present CLEO limits in units of $10^{-5}$.  
\label{fig:limits}}
\end{center}
\end{figure}

\noindent 
plane by
$\pm2.5 \sigma_M \times \pm 2.5 \sigma_{\Delta E}$ wide boxes, 
where $\sigma_M, \sigma_{\Delta E}$ denote $M_B$ and $\Delta E$ resolutions, respectively. 
Regions outside $3.5 \sigma_M$ and $\sigma_{\Delta E} $ are used for
background determination. No signal is observed in any of the 8 final states.
The observed yields are consistent with expected backgrounds.
The total efficiency is $4-5\%$ for modes containing a $K^0_S$ or a $K^{\ast+}$, $8-10\%$ 
for modes with a $K^{\ast0}$ and $17-23\%$ for modes with a $K^+$. Further details
are found in reference [17].

The resulting $90\%$ CL upper limits are plotted 
in Figure~\ref{fig:limits} together with those of previous studies, \cite{ca} recent 
CDF results \cite{alx} and a range of SM predictions.\cite{BABAR} 
The $K^{\ast 0} l^+l^-$ upper limit has been set for $m_{l^+l^-} > 0.5$~GeV, to 
reduce the contribution from $C^{(0)eff}_7$.
The new CDF and CLEO upper limits provide significant improvements, although
they still lie above the SM predictions. 
The CDF upper limit closest to the SM prediction is that for $ K^{\ast 0} \mu^+ \mu^-$ 
lying just a factor of 1.2 above. The most sensitive upper limit from CLEO is that
for the combined $K^{\ast 0} l^+l^-$ modes, which lies a factor of 2.2 above the SM predictions.

\section{Conclusion}

The present $B \ra X_s \gamma$ measurement is in good agreement with the NLO SM prediction.
The combined statistical and systematic errors are $15\%$, while the model dependence is
$8\%$. Even with the present accuracy,  $B \ra X_s \gamma$ already provides important
contraints on the SUSY parameter space. With the start of the BABAR, Belle
and CLEO III next year, the accuracy of the  $B \ra X_s \gamma$
branching fraction measurement will be significantly improved shortly.  
Sufficient $B$ mesons will be produced in these experiments to make an
observation of $B \ra X_s l^+ l^-$ modes possible.  
Thus, em penguin decays may actually provide the first hints on physics beyond the SM.

\section*{Acknowledgments}

I would like to acknowledge the CLEO collaboration for support with special thanks
to T. Skwarnicki and E.H. Thorndike. I would also like to thank J.L. Hewett for
sending me updated plots.


\section*{References}
\begin{thebibliography}{99}
\bibitem{bu}A.J. Buras {\it et al},  \Journal{\NPB}{424}{374}{1998};
A.Ali and C. Greub,\Journal{\ZPC} {60}{433}{1993}.

\bibitem{ch}K.G. Chetyrkin {\it et al},  \Journal{\PLB}{400}{206}{1997}.

\bibitem{BABAR}D.Boutigny {\it et al}, {\sl SLAC Report} 504, 1056pp (1998).

\bibitem{bu2}A.J. Buras {\it et al},  \Journal{\PLB}{414}{157}{1997}.

\bibitem{am}R. Ammar {\it et al}, \Journal{\PRL}{71}{674}{1993}.

\bibitem{al}M.S. Alam {\it et al}, \Journal{\PRL}{74}{2885}{1995}.

\bibitem{ali}A.Ali and C. Greub,   \Journal{\PLB}{259}{182}{1993}.

\bibitem{sj}T. Sj\"ostrand, Comp.Phys.Commun. 39, 347 (1986).

\bibitem{gl}S. Glenn  {\it et al}, preprint ICHEP98-1011, 11pp (1998).

\bibitem{ba}R.Barate  {\it et al},  \Journal{\PLB}{429}{169}{1998}.

\bibitem{he}J.L. Hewett and J.D. Wells,  \Journal{\PRD}{55}{5549}{1997}.

\bibitem{da}M.N.Danielsen  {\it et al}, proc. new dir. for HEP,
Snowmass Co. (1996).

\bibitem{he2}J.L. Hewett, private communication.

\bibitem{AM}A. Ali and T. Mannel {\it et al}, \Journal{\PLB}{264}{505}{1991}.

\bibitem{BBL}G. Buchalla {\it et al}, \Journal{Rev. Mod. Phys.}{68}{1125}{1996}.

\bibitem{gl2}S. Glenn  {\it et al}, \Journal{\PRL}{80}{2289}{1998}.

\bibitem{go}R. Godang  {\it et al}, preprint ICHEP98-1012, 18pp (1998).

\bibitem{ca}C. Caso {\it et al}, \Journal{\EPJC}{3}{1}{1998}.

\bibitem{alx}J. Alexander, plenary talk at ICHEP98, Vancouver, B.C., (1998). 

\end{thebibliography}
\end{document}